\PassOptionsToPackage{table,prologue}{xcolor}
\documentclass[sigconf]{acmart}
\AtBeginDocument{%
  }

\copyrightyear{2025}
\acmYear{2025}
\setcopyright{acmlicensed}
\acmConference[CIKM '25]{Proceedings of the 34th ACM International Conference on Information and Knowledge Management}{November 10--14, 2025}{Seoul, Republic of Korea}
\acmBooktitle{Proceedings of the 34th ACM International Conference on Information and Knowledge Management (CIKM '25), November 10--14, 2025, Seoul, Republic of Korea}
\acmDOI{10.1145/3746252.3761378}
\acmISBN{979-8-4007-2040-6/2025/11}

\definecolor{cloudblue}{rgb}{0.7294, 0.8, 0.851}
\definecolor{starblue}{rgb}{0.576, 0.710, 0.8118}

\usepackage{amsmath}
\usepackage{multirow}
\usepackage{booktabs}
\usepackage[normalem]{ulem}
\usepackage{enumitem}
\usepackage[most]{tcolorbox}
\usepackage{lipsum}    %
\usepackage{caption}
\usepackage{newfloat}

\definecolor{results}{RGB}{220, 230, 240}
\DeclareFloatingEnvironment{boxes} %
\newcommand{\boxref}[1]{\hyperref[{#1}]{Box~\ref*{#1}}}

\begin{document}

\title{SELF: Surrogate-light Feature Selection with Large Language Models in Deep Recommender Systems}

\author{Pengyue Jia}
\orcid{0000-0003-4712-3676}
\affiliation{%
  \institution{City University of Hong Kong}
  \city{Hong Kong SAR}
  \country{China}
}
\email{jia.pengyue@my.cityu.edu.hk}

\author{Zhaocheng Du}
\affiliation{
  \institution{Huawei Noah's Ark Lab}
  \city{Shenzhen}
  \country{China}}
\email{zhaochengdu@huawei.com}

\author{Yichao Wang}
\affiliation{
  \institution{Huawei Noah's Ark Lab}
  \city{Shenzhen}
  \country{China}}
\email{wangyichao5@huawei.com}

\author{Xiangyu Zhao}
\authornote{Corresponding author}
\affiliation{%
  \institution{City University of Hong Kong}
  \city{Hong Kong SAR}
  \country{China}
}
\email{xianzhao@cityu.edu.hk}

\author{Xiaopeng Li}
\affiliation{%
  \institution{City University of Hong Kong}
  \city{Hong Kong SAR}
  \country{China}
}
\email{xiaopli2-c@my.cityu.edu.hk}

\author{Yuhao Wang}
\affiliation{%
  \institution{City University of Hong Kong}
  \city{Hong Kong SAR}
  \country{China}
}
\email{yhwang25-c@my.cityu.edu.hk}

\author{Qidong Liu}
\affiliation{%
  \institution{City University of Hong Kong}
  \city{Hong Kong SAR}
  \country{China}
}
\email{qidongliu2-c@my.cityu.edu.hk}

\author{Huifeng Guo}
\affiliation{
  \institution{Huawei Noah's Ark Lab}
  \city{Shenzhen}
  \country{China}}
\email{huifeng.guo@huawei.com}

\author{Ruiming Tang}
\affiliation{
  \institution{Huawei Noah's Ark Lab}
  \city{Shenzhen}
  \country{China}}
\email{tangruiming@huawei.com}

\renewcommand{\shortauthors}{Jia et al.}

\begin{abstract}
Feature selection is crucial in recommender systems for improving model efficiency and predictive performance. 
Conventional approaches typically employ surrogate models—such as decision trees or neural networks—to estimate feature importance. However, their effectiveness is inherently constrained, as these models may struggle under suboptimal training conditions, including feature collinearity, high-dimensional sparsity, and insufficient data.
In this paper, we propose SELF, an \textbf{S}urrogat\textbf{E}-\textbf{L}ight \textbf{F}eature selection method for deep recommender systems. 
SELF integrates semantic reasoning from Large Language Models (LLMs) with task-specific learning from surrogate models. 
Specifically, LLMs first produce a semantically informed ranking of feature importance, which is subsequently refined by a surrogate model, effectively integrating general world knowledge with task-specific learning.
Comprehensive experiments on three public datasets from real-world recommender platforms validate the effectiveness of SELF.
\end{abstract}



\maketitle
\section{Introduction}

Feature selection~\cite{du2024tutorial,jia2024erase} is essential in deep recommender systems (DRS) to enhance model performance~\cite{wang2025tayfcs}, reduce overfitting, and accelerate both training and inference~\cite{wang2022autofield,lin2022adafs,du2024lightcs}. A variety of approaches have been proposed to address this need, which can be broadly classified into three categories based on their selection strategies:
(1) \textit{\textbf{Shallow feature selection methods}}~\cite{rf,xgboost}, which typically use statistical algorithms to assign importance scores to features. (2) \textit{\textbf{Gate-based feature selection methods}}~\cite{guo2022lpfs,lyu2023optfs} optimize a gate vector during training by assigning learnable gates to feature embeddings. The resulting gate values are interpreted as indicators of feature importance.
(3) \textit{\textbf{Sensitivity-based feature selection methods}}~\cite{wang2023sfs,Fperm,du2024lightcs} assess parameter sensitivity using gradient information obtained during backpropagation, and subsequently compute feature importance.

\begin{figure}[t]
    \centering
    \includegraphics[width=\linewidth]{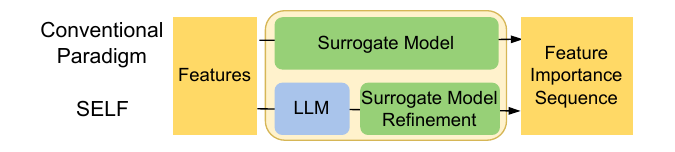}
    \caption{Comparison of conventional paradigm and SELF.}
    \label{fig:paradigm}
\end{figure}

\textbf{As illustrated in Figure~\ref{fig:paradigm}, all the aforementioned methods rely on training a surrogate model to approximate the feature-to-label mapping. The effectiveness of feature selection critically depends on how well this surrogate model reflects the ground truth.} However, in many real-world recommendation scenarios, surrogate models frequently fall short of delivering optimal performance. For instance, in cold-start or deep conversion tasks, the sparsity of samples—especially positive instances—often results in underfitting~\cite{schein2002methods}. In contrast, in scenarios characterized by numerous high-cardinality features, surrogate models are susceptible to overfitting~\cite{zhang2022towards}. Furthermore, these models commonly fail to capture interdependencies among features, thereby neglecting essential aspects such as collinearity and complementarity in the estimation of feature importance.

\begin{figure*}[ht]
    \centering
    \includegraphics[width=\textwidth]{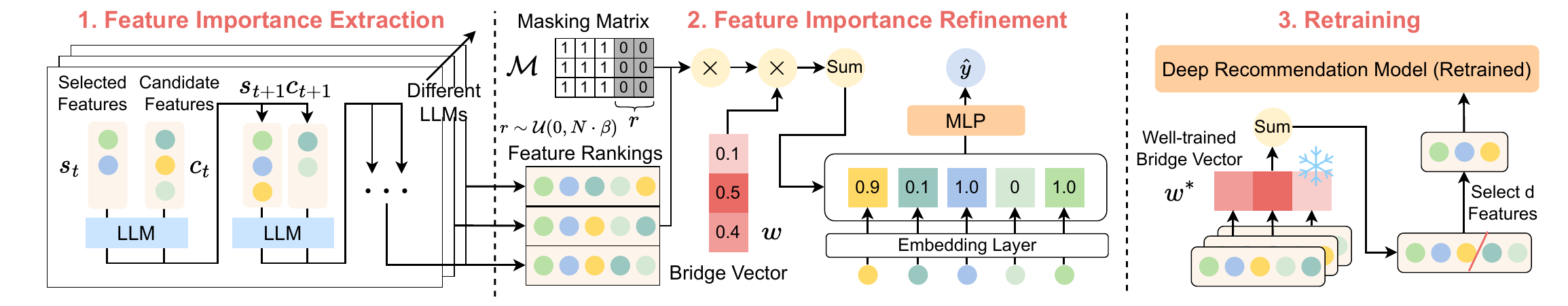}
    \caption{Overview of SELF. There are three stages in SELF, (1) Feature Importance Extraction Stage, (2) Feature Importance Refinement Stage, and (3) Retraining Stage.}
    \label{fig:overview}
\end{figure*}

LLMs offer promising solutions to the aforementioned longstanding challenges. Trained on vast corpora of web data, LLMs possess extensive world knowledge that enables them to identify informative feature subsets even in sparse data settings and to capture complex feature dependencies using semantic feature descriptions. For example, traditional surrogate models may fail to infer that ``longitude'' and ``latitude'' must appear together to uniquely define a geographic location, often selecting only one as part of the feature subset. In contrast, LLMs, by virtue of their pretrained knowledge, can inherently understand such relationships and thereby mitigate these limitations.

Despite these advantages, several key challenges must still be addressed to fully leverage LLMs in this context: (1) \textbf{Feature Complexity}: In real-world online platforms, many features exhibit intricate interdependencies, both with each other and with the target task. (2) \textbf{Knowledge Gap}: While LLMs are pretrained on general-domain knowledge, this may not be directly aligned with the domain-specific requirements of recommendation systems or other downstream tasks.
(3) \textbf{Efficiency Demand}: Moreover, industrial applications impose stringent constraints on computational efficiency and resource usage. Consequently, integrating feature importance derived from world knowledge into the training process of recommendation models in an efficient and lightweight manner remains a considerable challenge.

To address the challenges above, we propose SELF, a \textbf{S}urrogat\textbf{E}-\textbf{L}ight \textbf{F}eature selection method for DRS. Specifically, to explore complex feature interdependencies and feature-task relationships, we design a context-aware prompt iteration method that leverages LLMs to provide prior knowledge of feature importance. To efficiently and lightweightly integrate this knowledge into surrogate models for assessing feature importance, we introduce a bridge network to refine the feature importance rankings further. The bridge vector is trained under constraints to harmonize knowledge-based and task-specific feature importance. Our contributions can be summarized as follows:
\begin{itemize}
    \item We propose SELF, the first paradigm that integrates world knowledge priors with surrogate-based feature selection in DRS, effectively mitigating the inaccuracies of traditional methods that rely solely on the surrogate model.
    \item We develop a prompt iteration strategy that enables LLMs to iteratively select effective features by leveraging their world knowledge.
    \item We present a novel bridge network that aligns feature importance derived from world knowledge with the recommendation task space. This network is optimized in a lightweight, end-to-end manner.
    \item We conduct extensive experiments on three public datasets sourced from real-world platforms, demonstrating the effectiveness of SELF.
\end{itemize}

\section{Methodology}

In this section, we first provide an overview of the SELF framework in Section~\ref{sec:overview}. Next, we introduce the fundamental DRS model in Section~\ref{sec:DRS} and the key modules of SELF in Section~\ref{sec:stage1} and Section~\ref{sec:stage2}. Finally, we will describe the optimization and retraining in Section~\ref{sec:opt} and Section~\ref{sec:retrain}.

\subsection{Overview} \label{sec:overview}

In this subsection, we present an overview of \textbf{SELF}, which comprises three stages: (1) Feature Importance Extraction, (2) Feature Importance Refinement, and (3) Retraining, as illustrated in Figure~\ref{fig:overview}. In the Feature Importance Extraction stage, a prompt iteration strategy leverages LLMs to iteratively select predictive features from the candidate set, generating an initial feature importance ranking. To mitigate bias, multiple LLMs are employed in parallel. In the subsequent Feature Importance Refinement stage, the extracted importance sequences are refined via a bridge network integrated into the training process of the recommendation model. Finally, during the Retraining stage, the refined feature importance is fused through the optimized bridge vector, yielding a final ranking that incorporates semantic reasoning and aligns with the recommendation objective. The top-$d$ features from this ranking are then used to retrain the recommendation model from scratch.

\subsection{Basic DRS Architecture} \label{sec:DRS}

\subsubsection{Embedding Layer}
Efficient handling of categorical data is vital in recommender systems due to its inherent sparsity. Embedding layers address this challenge by transforming high-dimensional sparse categorical inputs into dense, low-dimensional vectors for DRS. The embedding process consists of two steps:  
(1) \textbf{Binarization}: Categorical features $\boldsymbol{x}$ are converted into binary vectors $\boldsymbol{x}'$ using one-hot encoding, where the dimensionality depends on the number of unique values in each feature field.  
(2) \textbf{Projection}: The binary vectors are projected into dense embeddings via embedding table lookups. For $n$-th feature field, the binary vector $\boldsymbol{x}_n'$ is mapped to an embedding $\boldsymbol{e}_n = \boldsymbol{x}_n' \cdot \boldsymbol{M}_n$, where $\boldsymbol{M}_n$ is the learnable embedding table, and $\boldsymbol{e}_n \in \mathbb{R}^{\text{dim}}$ is the resulting embedding. The concatenated output is $\boldsymbol{E} = [\boldsymbol{e}_1, \boldsymbol{e}_2, \cdots, \boldsymbol{e}_N]$, where $\boldsymbol{E}$ represents the combined feature embeddings.

\subsubsection{Transformation Layer}
The transformation layer is a key component in DRS, enabling robust fitting capabilities through linear transformations and nonlinear activation:
\begin{gather}
    \boldsymbol{h}_l = \text{ReLU}(\boldsymbol{W}_{l-1}\boldsymbol{h}_{l-1} + \boldsymbol{b}_{l-1}), \  \boldsymbol{h}_0 = \boldsymbol{E} \label{equ:trans-1}\\
    \hat{y} = \sigma(\boldsymbol{W}_{L}\boldsymbol{h}_{L}+\boldsymbol{b}_L) \label{equ:trans-2}
\end{gather}
where $\boldsymbol{h}_l$ is the hidden state of the $l$-th layer, and $\boldsymbol{W}_l$ and $\boldsymbol{b}_l$ are the weight and bias for the $l$-th layer. For the output layer, $\boldsymbol{h}_L$, $\boldsymbol{W}_L$, and $\boldsymbol{b}_L$ denote the hidden state, weight, and bias, respectively. $\sigma$ represents the activation function for recommendation tasks (e.g., Sigmoid and Softmax).

\subsection{Prompt Iteration} \label{sec:stage1}

To address the limitations of surrogate models in learning high-quality feature relationships, especially under conditions of feature sparsity or weak supervision, we introduce an external source of prior knowledge to guide the feature selection process. Specifically, we leverage the world knowledge embedded in LLMs to provide semantic priors that complement the surrogate model’s learning capacity. To this end, we design a prompt-based extraction strategy, as illustrated in the following box. The prompt comprises five components: instructions, descriptions, feature sets, supplementary information, and output formatting.

\begin{tcolorbox}[%
            colframe=starblue,
            width=1\linewidth,
            arc=1mm, 
            auto outer arc,
            title={Prompt structure},
            label={box:prompt}]
    \textbf{[Instructions]}
    
    You are a professional researcher in recommender systems and feature selection. ...
    
    \textbf{[Descriptions]} 
    
    Task: \textless Task descriptions \textgreater
    
    Dataset: \textless Dataset descriptions \textgreater
    
    Features: \textless Feature descriptions \textgreater

    \textbf{[Feature sets]}
    
    Features already selected: 
    
    \textless A set contains all selected features \textgreater

    Candidate features: 
    
    \textless A set contains all candidate features \textgreater

    \textbf{[Supplementary information]}
    
    Please select one feature from the candidate feature set that you think is most important for this task to add to the selected features. Suppose we will discrete the features based on their unique values. 
    
    The feature you choose should, as much as possible, have the following characteristics: 

    1. They are informative.
    
    2. Independent of other selected features. 
    
    3. Simple and easy for the model to understand.

    \textbf{[Output formatting]}
    
    Your answer's last line should be the feature name of your selected feature, without any other characters.
    \end{tcolorbox}
\noindent(1) \textbf{Instructions} defines the role of LLMs and introduces the subsequent tasks. 
(2) \textbf{Descriptions} provide critical information for addressing the feature selection problem, encompassing the recommendation task, dataset, and feature set. For the task description, we outline the context and objectives of the recommendation scenario.
For instance, in the MovieLens dataset, the task involves predicting whether a user will like a movie based on user profiles and movie information. The dataset description focuses on the number of samples and features. The feature set is categorized into user, item, and interaction features. Each feature is described using five attributes: name, description, data type (e.g., integer, string), example value, and the number of unique values. This structured representation offers sufficient context for LLMs to assess feature importance effectively.
(3) \textbf{Feature Sets} are composed of two parts: features already selected and candidate features. The union of these two sets constitutes the complete feature set. (4) \textbf{Supplementary Information}  emphasizes the expected actions of LLMs (selecting one feature from the candidate features to add to the already selected features during each inference) and provides reference criteria for feature selection. (5) \textbf{Output Formatting} standardizes the output format of LLMs to facilitate automated and streamlined feature selection. 
Inspired by existing work~\cite{sun2023chatgpt}, we prompt LLMs to analyze and select informative features in an iterative manner. Considering that different LLMs may exhibit significant assessing bias due to variations in training data, we treat the LLM as an expert while simultaneously introducing multiple different LLMs to generate several feature importance sequences:
\begin{gather}
    f_{t}^{k} = \text{LLM}_k(\text{prompt}, \boldsymbol{s}_{t}^{k}, \boldsymbol{c}_{t}^{k}) \\
    \boldsymbol{s}_{t+1}^{k} = \boldsymbol{s}_{t}^{k} \cup \{f_{t}^{k}\}, \ 
    \boldsymbol{c}_{t+1}^{k} = \boldsymbol{c}_{t}^{k} \setminus \{f_{t}^{k}\} \label{eqa:equation2}
\end{gather}
where $\boldsymbol{s}_t^{k}$ and $\boldsymbol{c}_t^{k}$ denote the sets of selected and candidate features, respectively, at the $t$-th step for the $k$-th LLM, and $f_t^k$ represents the feature selected at that step. The selected feature $f_t^k$ is then used to update the selected and candidate sets according to Equation~\ref{eqa:equation2}.
Given that user ID and item ID are fundamental in recommendation tasks, we initialize $\boldsymbol{s}_0$ as a set containing these two features.
The iterative process continues until the candidate feature set becomes empty. The final output is a matrix that records the features selected at each time step by different LLMs:
\begin{equation}
    \boldsymbol{\mathcal{S}} = 
        \begin{bmatrix}
        f_0^{0} & f_1^{0} & \cdots & f_N^{0} \\
        f_0^{1} & f_1^{1} & \cdots & f_N^{1} \\
        \vdots & \vdots & \ddots & \vdots \\
        f_0^{K} & f_1^{K} & \cdots & f_N^{K} \\
        \end{bmatrix}
\end{equation}
where each row corresponds to the feature selection trajectory of a specific LLM, and each column indicates the feature selected at a particular iteration step. Here, $K$ denotes the number of LLMs, and $N$ is the total number of selection steps.

\subsection{Bridge Network} \label{sec:stage2}
The feature importance rankings obtained via prompting LLMs rely exclusively on general world knowledge, which may diverge from the task-specific importance signals learned during model training. This discrepancy arises from a distributional mismatch between knowledge priors and empirical data. To effectively integrate knowledge-derived importance into the surrogate model's learning process, we introduce a bridge network that acts as an intermediary, aligning the semantic priors from LLMs with the data-driven objectives of the recommendation model.

For each training batch, we mask a subset of features according to the feature importance rankings produced by LLMs, beginning with the least important ones. The number of masked features, denoted by $r$, is sampled from a uniform distribution: $r \sim \mathcal{U}(0, N \cdot \beta)$, where $N$ is the total number of feature fields and $\beta$ is the maximum masking ratio.
\begin{gather}
    \boldsymbol{\mathcal{M}} = 
        \begin{bmatrix}
        m_0^{0} & \cdots & m_N^{0} \\
        m_0^{1} & \cdots & m_N^{1} \\
        \vdots & \ddots & \vdots \\
        m_0^{K} & \cdots & m_N^{K} \\
        \end{bmatrix}, 
        \ m_{t}^{k} = \begin{cases}
            0&\text{if} \ t \geq N-r \\
            1&\text{otherwise}
        \end{cases} \label{equ:masking}
\end{gather}
In this formulation, $\boldsymbol{\mathcal{M}}$ denotes the masking matrix, where $m_t^k$ indicates the masking value at the $t$-th selection step for the $k$-th LLM. A value of $m_t^k = 0$ means the corresponding feature is masked, while $m_t^k = 1$ indicates that the feature is retained for training.

To incorporate the feature importance rankings from multiple LLMs into the training process, we introduce a learnable \textit{bridge vector} that adaptively fuses their contributions. Let the feature set be denoted by $F$, and let $\boldsymbol{w} = [w_0, w_1, \dots, w_K]$ represent the bridge vector, where $K$ is the number of LLMs. The fusion weights are normalized via a temperature-scaled softmax:
\begin{equation}
    \boldsymbol{\hat{w}} = \text{Softmax}(\boldsymbol{w} \cdot \exp(\tau))
\end{equation}
where $\tau$ is the temperature parameter controlling the sharpness of the softmax distribution.

For each feature field $\boldsymbol{F}_n$, we compute its final importance score $\boldsymbol{h}_n$ as follows:
\begin{equation}
    \boldsymbol{h}_n = \sum_{k=0}^{K} \hat{w}_k \cdot \sum_{t=0}^{N} \mathbb{I}[f_t^k = \boldsymbol{F}_n] \cdot m_t^k
\end{equation}

\noindent where $\mathbb{I}[\cdot]$ is the indicator function that evaluates to 1 if feature $\boldsymbol{F}_n$ appears at step $t$ in the ranking of the $k$-th LLM, and 0 otherwise. $m_t^k$ is the masking value defined in Equation~\ref{equ:masking}, indicating whether the feature at step $t$ in the $k$-th LLM’s sequence is retained. $\hat{w}_k$ denotes the normalized importance weight assigned to the $k$-th LLM. This formulation enables the model to dynamically weigh and aggregate unmasked feature importance across different LLM-generated rankings, producing a consolidated importance score $\boldsymbol{h}_n$ for each feature field.

Before applying the integrated weights $\boldsymbol{h}$ to the feature embeddings, we first normalize the embeddings to ensure consistency across feature fields. 
After transformation, the embeddings of features should be $\boldsymbol{\hat{E}}=[\boldsymbol{\hat{e}}_1, \boldsymbol{\hat{e}}_2, \cdots, \boldsymbol{\hat{e}}_N]$. Next, we weight embeddings by $\boldsymbol{E}^{\prime} = [\boldsymbol{h}_1 \boldsymbol{\hat{e}}_1, \boldsymbol{h}_2 \boldsymbol{\hat{e}}_2, \cdots, \boldsymbol{h}_N \boldsymbol{\hat{e}}_N]$,
where $\boldsymbol{E}^{\prime}$ denotes weighted embeddings, $\boldsymbol{h}_n$ and $\boldsymbol{\hat{e}}_n$ are the weight and embeddings after normalization for $n$-th feature field. The weighted embeddings will be input into the transformation layer in DRS to generate predictions following Equation~\ref{equ:trans-1} and Equation~\ref{equ:trans-2}.

\subsection{Optimization} \label{sec:opt}

We take the Click-through rate (CTR) prediction as our task in this paper, and we use the binary cross-entropy (BCE) loss as our optimization objective:
\begin{equation}
\text{min} \ \mathcal{L} = -(\boldsymbol{y}_i \text{log}(\hat{\boldsymbol{y}}_i) + (1-\boldsymbol{y}_i)\text{log}(1-\hat{\boldsymbol{y}}_i))
\end{equation}
where $\boldsymbol{y}$ and $\hat{\boldsymbol{y}}$ are the ground truth label and the prediction value.

\subsection{Retraining} \label{sec:retrain}

In this subsection, we describe how to select informative feature fields based on the learned fusion weights and retrain the deep recommender system.

Given the well-trained bridge vector $\boldsymbol{w}^{*}$, we compute the final fusion weights $\hat{\boldsymbol{w}}^{*} = \text{Softmax}(\boldsymbol{w}^{*}\cdot \exp(\tau))$, which reflect the reliability of each LLM expert. For each feature ranked at position $t$ by the $k$-th LLM, we assign a linearly decaying importance score:
\begin{equation}
\text{fi}_t^k = 1 - \frac{t}{N}
\end{equation}
where $\text{fi}_t^k$ denotes the relative importance of the $t$-th ranked feature in the $k$-th LLM ranking.
We then aggregate the importance scores across all experts to obtain the final importance score $\boldsymbol{h}_{n}^{*}$ for the $n$-th feature field:
\begin{equation}
\boldsymbol{h}_n^{*} = \sum_{k=0}^{K} \hat{w}_k^* \cdot \sum_{t=0}^{N} \mathbb{I}[f_t^k = \boldsymbol{F}_n] \cdot \text{fi}_t^k
\end{equation}
where
$\hat{w}_k^{*}$ is the learned fusion weight for the $k$-th LLM. $\mathbb{I}[f_t^k = \boldsymbol{F}_n]$ is the indicator function that equals 1 if the $n$-th feature field appears at position $t$ in the $k$-th ranking. $\boldsymbol{h}_n^*$ is the final integrated importance score for feature field $\boldsymbol{F}_n$.

We then select the top-$d$ feature fields with the highest $\boldsymbol{h}_n^*$ values and retrain the deep recommender system from scratch using only the selected features.

\section{Experiments}

To achieve a comprehensive understanding of SELF, we aim to answer the following research questions:
\begin{itemize}[leftmargin=*]
    \item \textbf{RQ1:} How does the performance of SELF compare to the other state-of-the-art feature selection methods in DRS?
    \item \textbf{RQ2:} How does the transferability of SELF to other backbone deep recommendation models?
    \item \textbf{RQ3:} What's the specific effect of the proposed components?
    \item \textbf{RQ4:} How SELF performs under data-scarcity scenarios?
    \item \textbf{RQ5:} How do hyperparameters influence performance?
    \item \textbf{RQ6:} How SELF performs in real industrial environments?
\end{itemize}

\subsection{Experimental Settings}

\begin{table}[h]
\centering
\caption{Dataset statistics}
\label{tab:app_dataset}
\resizebox{0.9\linewidth}{!}{
\begin{tabular}{cccc} 
\toprule
Dataset          & Movielens-1M & Aliccp     & Kuairand   \\ 
\midrule
Interactions     & 1,000,209    & 85,316,519 & 1,436,609  \\
Users            & 6,040        & 238,635    & 27,285     \\
Items            & 3,706        & 467,298    & 7,551      \\
Interaction Type & Rating (1-5) & Click      & Click      \\
Feature Num      & 9            & 23         & 96         \\
\bottomrule
\end{tabular}}
\end{table}

\begin{table*}
\setlength{\tabcolsep}{16pt}
\centering
\caption{Comparison results between SELF and other state-of-the-art baselines on three public datasets. The metrics of the best-performed methods and sub-optimal methods are highlighted in \textbf{bold} fonts and \underline{underlined}. 
$\uparrow$ denotes the higher is better. $\downarrow$ denotes the lower is better.}
\label{tab:overall_performance}
\begin{tabular}{ccccccc}
\toprule
\multirow{2}{*}{Dataset} & \multicolumn{2}{c}{Movielens-1M}        & \multicolumn{2}{c}{Aliccp}              & \multicolumn{2}{c}{Kuairand}             \\ 
\cline{2-7}
                         & AUC $\uparrow$   & Logloss $\downarrow$ & AUC $\uparrow$   & Logloss $\downarrow$ & AUC $\uparrow$   & Logloss $\downarrow$  \\ 
\hline
No Selection             & 0.78854          & 0.54367              & 0.61804          & 0.16190              & 0.77953          & 0.55872               \\ 
\hline
\multicolumn{7}{c}{\textit{Shallow Feature Selection}}                                                                                                  \\ 
\hline
Lasso                    & 0.78853          & 0.54388              & 0.61861          & 0.16185              & 0.77987          & 0.55835               \\
GBDT                     & 0.78904          & 0.54318              & 0.61854          & 0.16232              & 0.77972          & 0.55835               \\
RF                       & 0.78868          & 0.54336              & 0.61910          & 0.16172              & 0.77988          & \textbf{0.55801}      \\
XGBoost                  & 0.80081          & 0.53075              & 0.61873          & 0.16176              & 0.77992          & \uline{0.55806}       \\ 
\hline
\multicolumn{7}{c}{\textit{Gate-based Feature Selection}}                                                                                               \\ 
\hline
AutoField                & 0.80315          & 0.53075              & 0.61922          & \uline{0.16141}      & \uline{0.77997}  & 0.55820               \\
AdaFS                    & 0.78523          & 0.54765              & 0.61869          & 0.16143              & 0.77981          & 0.55854               \\
OptFS                    & 0.77606          & 0.55564              & 0.61862          & 0.16264              & 0.77933          & 0.55855               \\
LPFS                     & \uline{0.80339}  & \uline{0.52822}      & 0.61836          & 0.16189              & 0.77983          & 0.55824               \\ 
\hline
\multicolumn{7}{c}{\textit{Sensitivity-based Feature Selection}}                                                                                        \\ 
\hline
Permutation              & 0.78853          & 0.54354              & \uline{0.61935}  & 0.16152              & 0.77978          & 0.55831               \\
SFS                      & 0.78927          & 0.54241              & 0.61897          & 0.16180              & 0.77981          & 0.55819               \\
SHARK                    & 0.78904          & 0.54277              & 0.61928          & 0.16151              & 0.77995          & 0.55815               \\ 
\hline
\rowcolor{results} SELF                     & \textbf{0.80480} & \textbf{0.52703}     & \textbf{0.62015} & \textbf{0.16139}     & \textbf{0.78002} & \textbf{0.55801}      \\
\bottomrule
\end{tabular}
\end{table*}

\subsubsection{\textbf{Dataset.}} We conduct experiments on three public datasets: Movielens-1M\footnote{https://grouplens.org/datasets/movielens/1m/}, Aliccp~\cite{ma2018entire}, and Kuairand~\cite{gao2022kuairand}. These datasets have varying numbers of features (9 in Movielens-1M, 23 in Aliccp, and 96 in Kuairand). They are chosen to illustrate the effectiveness of our method under different application conditions. The statistics of datasets are illustrated in Table~\ref{tab:app_dataset} and the detailed descriptions of the datasets are given in \textbf{Appendix~\ref{sec:app_dataset}}.

\subsubsection{\textbf{Baselines.}} We perform experiments with the following baselines to demonstrate the advanced performance of SELF. 

\noindent\textbf{Shallow Feature Selection Methods:}
\begin{itemize}[leftmargin=*]
    \item \textbf{Lasso~\cite{lasso}.} Lasso is the Least Absolute Shrinkage and Selection Operator, which effectively combines variable selection and regularization to enhance model performance.
    \item \textbf{GBDT~\cite{gbdt}.} Gradient Boosted Decision Trees (GBDT) is a highly effective machine learning algorithm that leverages the strengths of decision trees in combination with the boosting technique. It constructs an ensemble of decision trees, with each tree iteratively correcting the errors of its predecessors.
    \item \textbf{Random Forest~\cite{rf}.} Random Forest determines feature importance by measuring the extent to which each feature reduces impurity within a decision tree.
    \item \textbf{XGBoost~\cite{xgboost}.} XGBoost is a highly efficient and scalable machine learning algorithm built on the principles of gradient boosting. XGBoost determines feature importance by assessing the impact of each feature on the model's final predictive performance.
\end{itemize}

\noindent\textbf{Gate-based Feature Selection Methods:}
\begin{itemize}[leftmargin=*]
    \item \textbf{AutoField~\cite{wang2022autofield}.} AutoField is the first work in the recommender systems domain to focus on feature selection within DRS (Deep Recommender Systems). It introduces an advanced controller network that trains two parameters for each feature field, representing the likelihood of selecting that feature.
    \item \textbf{AdaFS~\cite{lin2022adafs}.} AdaFS refines feature selection further along the lines of AutoField. Recognizing that feature importance varies significantly across different samples, AdaFS focuses on generating dynamic gating weights for each sample, enabling the model to adaptively select features.
    \item \textbf{OptFS~\cite{lyu2023optfs}.} OptFS differs from previous work by focusing on feature value selection. Through the learning of a gating network, OptFS can significantly reduce the number of feature values while simultaneously enhancing the  performance.
    \item \textbf{LPFS~\cite{guo2022lpfs}.} LPFS critiques previous approaches that determine feature importance based on the magnitude of gating values. It introduces a smoothed-$l^0$
    function that effectively selects features in a single stage.
\end{itemize}

\noindent\textbf{Sensitivity-based Feature Selection Methods:}
\begin{itemize}[leftmargin=*]
    \item \textbf{Permutation~\cite{fisher2019allpermutation}.} This method assesses the importance of a feature by disrupting the values within a specific feature field and observing whether this permutation significantly impacts the model's predictive performance.
    \item \textbf{SFS~\cite{wang2023sfs}.} SFS assigns feature importance weights by examining the gradient of the gating vector on a small subset of data.
    \item \textbf{SHARK~\cite{Fperm}.} SHARK uses the first-order component of the Taylor expansion as a measure of feature importance. It improves model efficiency and performance by removing less important features from the embedding table.
\end{itemize}

\subsubsection{\textbf{Evaluation Metrics.}} To comprehensively evaluate SELF, we use the AUC and Logloss metrics in CTR prediction following previous work~\cite{wang2022autofield, lin2022adafs}. It is worth noting that an increase of 0.001 in AUC or a decrease in Logloss is already a significant improvement in the CTR prediction task~\cite{song2019autoint,wang2021dcn}.

\subsubsection{\textbf{Implementation Details.}} 

We use Adam~\cite{kingma2014adam} optimizer with learning rate 0.001, $\beta_1=0.9$, $\beta_2=0.999$, and $\epsilon=1 \times 10^{-8}$ for all experiments. 
We set the batch size to 4096 and the embedding dimension to 8. The number of LLMs is $K = 3$ (GPT4, GPT4o, and GPT-3.5), the maximum masking ratio is 0.2, and the temperature coefficient is $\tau = 4.0$. To mitigate the effect of randomness, each experiment is repeated three times, and the average performance is reported. Detailed implementation settings are provided in \textbf{Appendix~\ref{sec:app_imp}}.

\subsection{Overall Performance (RQ1)}

In this section, we compare SELF with other state-of-the-art baselines on three public datasets. The results are listed in Table~\ref{tab:overall_performance}, and we can derive the following findings: 
1) SELF achieves the best performance across all metrics on three datasets. This superior performance is attributed to SELF's unique approach, which does not rely solely on the information provided by surrogate models to determine feature importance. Instead, it combines both world knowledge and task-specific information to comprehensively assess feature importance. 2) Overall, Gate-based and Sensitivity-based feature selection methods outperform Shallow feature selection methods. This is because Shallow feature selection methods rely on the original classifier, whose performance is generally weaker compared to the deep neural networks employed by Gate-based and Sensitivity-based feature selection methods. 3) The performance improvement of SELF compared to no selection varies significantly across different datasets. Specifically, the improvement is most pronounced on the Movielens-1M dataset, while it is relatively modest on the Kuairand dataset. This discrepancy is due to the substantial differences in feature quantities among the datasets. The Movielens-1M dataset has fewer features, so removing a certain amount of irrelevant features has a more significant impact and yields greater gains. 

\subsection{Transferability Study (RQ2)}

In this section, we investigate the transferability of features selected by SELF. Specifically, whether the selected features remain effective when applied to other deep recommendation backbone models. We apply the features selected by SELF to four popular deep recommendation models: FM~\cite{rendle2010factorization}, DeepFM~\cite{guo2017deepfm}, Wide\&Deep~\cite{cheng2016wide}, and DCN~\cite{wang2017deep} on the Aliccp dataset. In Table~\ref{tab:transferability}, ``No Selection'' and ``SELF'' represent the performance of the model with all available features and with the features selected by SELF. We can find that: 1) All deep recommendation backbone models achieve improved prediction performance after applying the features selected by SELF. This demonstrates that the features selected by SELF possess strong transferability capabilities and can adapt to various deep recommender systems. 2) By comparing the performance improvements across different models after applying the SELF-selected features, we observe that more complex models (Wide\&Deep and DCN) show greater enhancements compared to FM-based models. The possible reason is that more powerful models tend to amplify the impact of feature selection. 

\begin{table}[t]
\centering
\caption{Transferability study.}
\label{tab:transferability}
\resizebox{0.9\linewidth}{!}{
\begin{tabular}{c|cccc} 
\toprule
\multirow{2}{*}{Model} & \multicolumn{2}{c}{AUC~$\uparrow$} & \multicolumn{2}{c}{Logloss~$\downarrow$}  \\
                       & No Selection & SELF               & No Selection & SELF                      \\ 
\midrule
FM                     & 0.60667      & \textbf{0.60725}    & 0.16284      & \textbf{0.16283}           \\
DeepFM                 & 0.61716      & \textbf{0.61757}    & 0.16203      & \textbf{0.16180}           \\
WideDeep               & 0.62084      & \textbf{0.62212}    & 0.16143      & \textbf{0.16120}           \\
DCN                    & 0.62271      & \textbf{0.62322}    & 0.16173      & \textbf{0.16124}           \\
\bottomrule
\end{tabular}}
\end{table}

\subsection{Ablation Study (RQ3)}

\begin{figure}[t]
    \centering
    \includegraphics[width=0.98\linewidth]{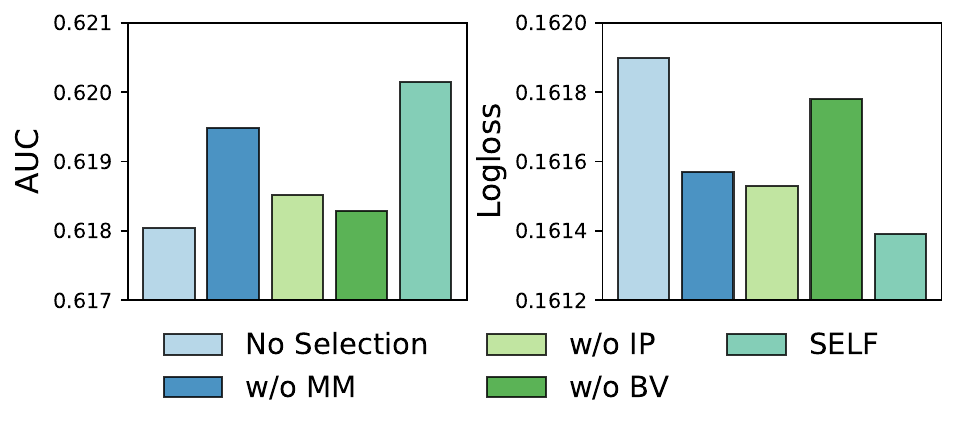}
    \caption{Ablation Study.}
    \label{fig:ablation_study}
\end{figure}

\begin{table*}
\setlength{\tabcolsep}{16pt}
\centering
\caption{Data-scarcity performance of SELF and baselines on three datasets. The metrics of best-performed methods are highlighted with \textbf{bold} fonts, and the sub-optimal methods are emphasized with \underline{underlined} text.}
\label{tab:scarce_performance}
\begin{tabular}{c|cccccc} 
\toprule
\multirow{2}{*}{Dataset} & \multicolumn{2}{c}{Movielens-1M}        & \multicolumn{2}{c}{Aliccp}              & \multicolumn{2}{c}{Kuairand}             \\ 
\cline{2-7}
                         & AUC~$\uparrow$   & Logloss~$\downarrow$ & AUC~$\uparrow$   & Logloss~$\downarrow$ & AUC~$\uparrow$   & Logloss~$\downarrow$  \\ 
\hline
No Selection             & 0.70482          & 0.68468              & 0.55555          & 0.16867              & 0.70704          & 0.67961               \\ 
\hline
\multicolumn{7}{c}{\textit{Shallow Feature Selection}}                                                                                                  \\ 
\hline
Lasso                    & 0.71338          & 0.68496              & 0.55634          & 0.16812              & 0.71161          & 0.68816               \\
GBDT                     & 0.70131          & 0.66595              & \uline{0.55837}  & 0.16800              & 0.71398          & 0.67449               \\
RF                       & 0.70416          & 0.68539              & 0.55639          & 0.16846              & 0.71308          & 0.67291               \\
XGBoost                  & 0.71737          & 0.62630              & 0.55636          & 0.16865              & \uline{0.71418}  & 0.67441               \\ 
\hline
\multicolumn{7}{c}{\textit{Gate-based Feature Selection }}                                                                                              \\ 
\hline
AutoField                & 0.72051          & \textbf{0.62377}     & 0.55823          & 0.16818              & 0.71256          & 0.67837               \\
AdaFS                    & 0.69554          & 0.65557              & 0.55222          & \textbf{0.16666}     & 0.70571          & 0.67178               \\
OptFS                    & 0.69641          & 0.66066              & 0.55610          & 0.16926              & 0.70682          & \textbf{0.65146}      \\
LPFS                     & 0.70213          & 0.68940              & 0.55768          & 0.16767              & 0.71292          & 0.68397               \\ 
\hline
\multicolumn{7}{c}{\textit{Sensitivity-based Feature Selection }}                                                                                       \\
\hline
Permutation              & 0.70700          & 0.69110              & 0.55749          & 0.16785              & 0.71370          & 0.67278               \\
SFS                      & \uline{0.72232}  & 0.64441              & 0.55570          & 0.16781              & 0.71239          & 0.68463               \\
SHARK                    & 0.70318          & 0.68607              & 0.55758          & 0.16798              & 0.71359          & 0.68180               \\ 
\hline
\rowcolor{results}
 SELF                     & \textbf{0.72372} & \uline{0.62563}      & \textbf{0.55878} & \uline{0.16735}      & \textbf{0.71463} & \uline{0.66995}       \\
\bottomrule
\end{tabular}
\end{table*}

In this section, we compare the performance of the following variants on Aliccp to verify the effectiveness of each component: 1) \textbf{w/o MM}: SELF without multiple LLMs. This variant only considers the feature sequence generated by one advanced LLM (GPT4\footnote{\url{https://openai.com/index/gpt-4/}}). 2) \textbf{w/o IP}: SELF without iterative prompt. This variant prompts the LLM to generate feature sequences in one shot, not in an iterative manner. 3) \textbf{w/o BV}: SELF without bridge vector. This variant directly derives the final feature rankings by averaging the feature sequences generated by LLMs.
From Figure~\ref{fig:ablation_study}, we can observe the following conclusions: 1) SELF outperforms w/o MM. It is because the feature importance ranking from a single advanced LLM has an inherent bias, while SELF can combine rankings from multiple LLMs to achieve a more robust and effective ranking. 2) SELF is better than w/o IP. The possible reason is that the iterative prompt method for obtaining feature importance rankings allows the LLM to focus on a more straightforward task each time, resulting in better performance than the one-shot approach. 3) SELF is superior to w/o BV, indicating the effectiveness of the bridge vector, which can effectively evaluate the correctness of different feature rankings and integrate them. 4) All variants outperform ``No Selection'' in all metrics, demonstrating the stability of SELF.

\subsection{Data-scarcity Performance (RQ4)}

In this section, we discuss the performance of SELF under data-scarcity conditions. To simulate a data-limited scenario, we reduce the training and validation sets of public datasets to 5\% of their original size. Table~\ref{tab:scarce_performance} presents the results of SELF and the baselines across three datasets. Based on Table~\ref{tab:scarce_performance}, we can draw the following conclusions: 1) SELF maintains its stable superiority even in data-scarce scenarios, almost outperforming all baselines across all metrics. This stability is due to SELF's ability to optimize the bridge vector with a minimal amount of data, thereby enhancing its effectiveness under limited data conditions. 2) Compared to the results in Table~\ref{tab:overall_performance}, the gap between SELF and the No Selection approach becomes more pronounced. This indicates that feature selection has a more significant impact on the final results when data is scarce. 3) In the MovieLens-1M dataset, six baselines perform even worse than the No Selection approach. This is likely because, under data-scarce conditions, models are prone to overfitting, which prevents the surrogate model from providing reliable information for assessing feature importance.

\subsection{Hyperparameter Analysis (RQ5)}

\begin{figure*}
    \centering
    \includegraphics[width=\textwidth]{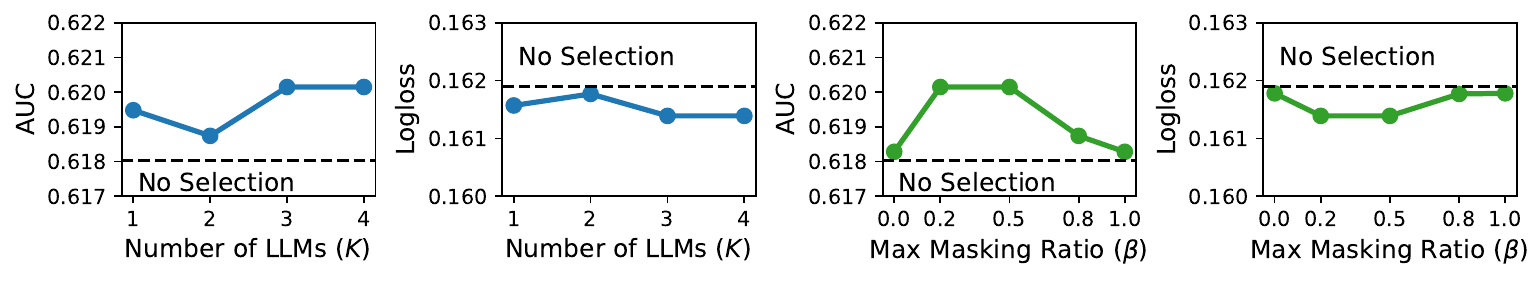}
    \caption{Hyperparameter analysis on number of LLMs ($K$) and maximum masking ratio ($\beta$).}
    \label{fig:hyperparameter}
\end{figure*}

In this section, we will investigate the effect of hyperparameters on the performance of the Aliccp dataset.
There are two important hyperparameters in SELF: the number of LLMs ($K$) and the maximum masking ratio ($\beta$). Figure~\ref{fig:hyperparameter} shows the results of AUC and Logloss on different hyperparameters. 

\subsubsection{Number of LLMs.} As shown in the left portion of Figure~\ref{fig:hyperparameter}, we experiment with the number of LLMs from 1 to 4 (GPT4, GPT4o, GPT3.5, and Gemini-1.5-Pro). We prioritize using models with larger parameter sizes, starting with the most powerful model. The results lead to the following conclusions: (1) A single strong LLM can provide valuable information for feature selection, with just one LLM outperforming the No selection baseline. (2) The optimal performance is achieved when the number of LLMs is set to 3. This is because the bridge vector can effectively assign weights to the existing feature importance sequences generated by different LLMs according to the current recommendation task. This results in a comprehensive feature importance ranking that integrates both semantic and task-specific aspects. (3) When the number of LLMs exceeds 3, the performance of the method stabilizes, indicating the generalization and stability of SELF, with 3 being the most cost-effective hyperparameter choice.

\subsubsection{Maximum Masking Ratio.} As illustrated in the right portion of Figure~\ref{fig:hyperparameter}, we set the maximum masking ratio from 0 to 1 and freeze other hyperparameters. We can draw the following conclusions: 1) The overall AUC trend exhibits an initial increase followed by a decrease. This is because when $\beta$ approaches 0, no features will be masked, resulting in the final feature ranking being a simple average of the sequences generated by LLMs. Conversely, when $\beta$ approaches 1, the most important features in the sequences generated by the LLMs tend to be similar at the top, while the differences among the features in the lower ranks are gradually diminished as beta increases. This weakens the effectiveness of feature selection. 2) We can also observe that the feature selection performance remains consistently exceptional when $\beta$ is between 0.2 and 0.5, highlighting the robustness and generalization capability of SELF.

\subsection{Online Experiments (RQ6)}

We deployed SELF on a large-scale industrial search advertising platform that serves billions of users daily. Users interact with the platform by submitting search queries and receiving responses about relevant applications and associated advertisements. In our system, ads from different advertisers are divided into two stages based on their onboarding time: a learning phase and a stable phase. Ads in the learning phase typically suffer from limited user feedback due to their short exposure duration, leading to low-quality and sparse training samples. Under such conditions, traditional data-driven feature selection methods—such as AutoField and SFS—struggle to identify high-quality features, as they heavily depend on the availability of abundant, high-quality labeled data.

To address the challenges in the learning phase, we deployed the SELF algorithm to assist with feature selection under limited data conditions. Specifically, we provided the Qwen family models with business-related features and their corresponding descriptions, along with a clear explanation of the recommendation task. This enabled the LLMs to generate initial feature importance rankings based on semantic reasoning. We then refine these rankings using the limited training data available in the learning phase. Based on the resulting importance scores, we removed the 12 least important features and deployed the resulting model as the experimental group. All other parameters were held constant, isolating the impact of feature selection. Compared to the baseline model, the experimental model achieved a \textbf{\uline{0.63\%}} improvement in RPM (Revenue per Mille), a \textbf{\uline{3.01\%}} increase in CTR (Click-Through Rate), and a \textbf{\uline{13.6\%}} reduction in online inference latency. This optimized feature set has since been adopted to serve the entire production traffic.

\subsection{Case Study} \label{sec:appendix_case}

\begin{figure}
    \centering
    \includegraphics[width=\linewidth]{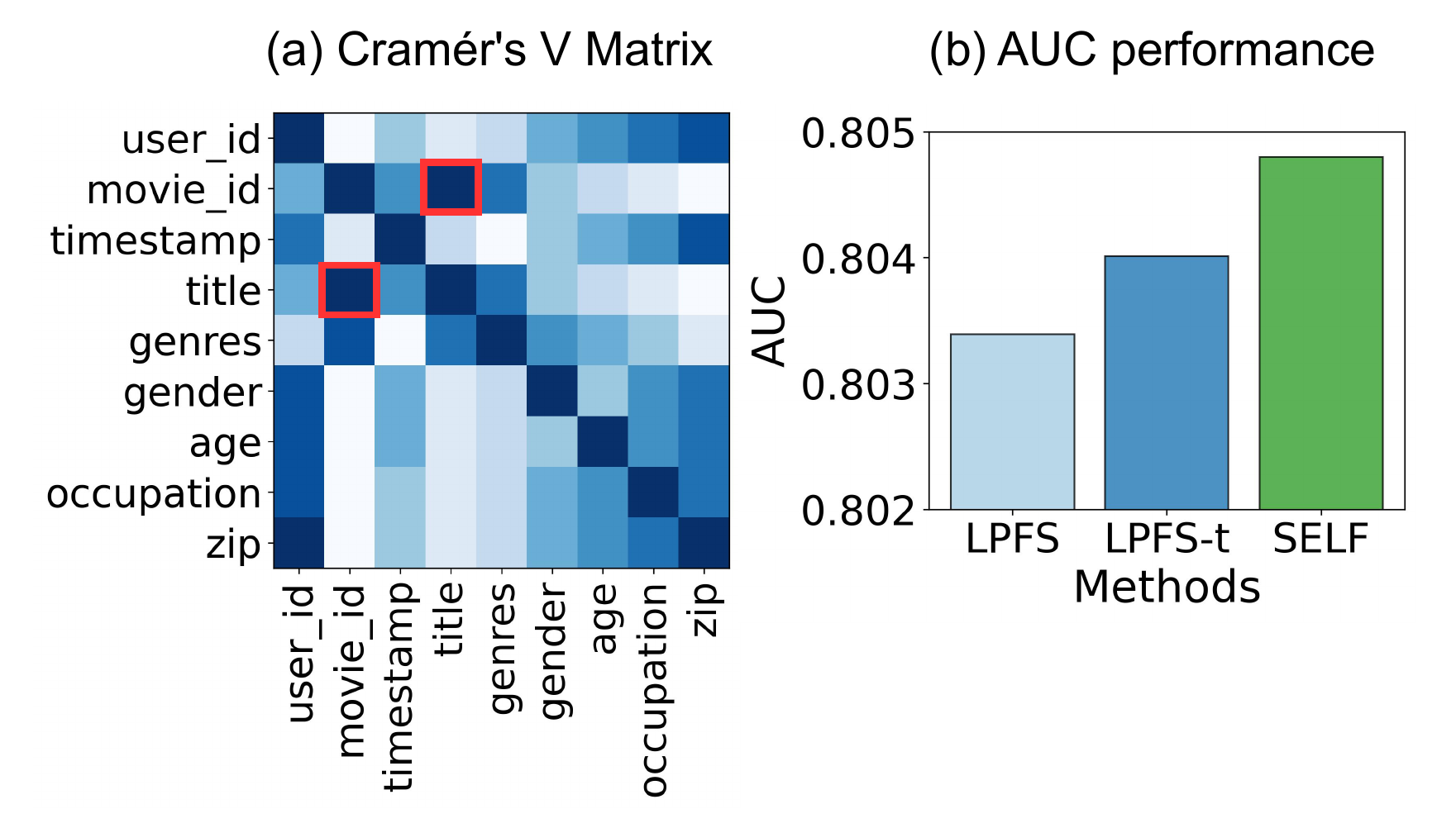}
    \caption{(a) Cramér's V matrix and (b) AUC performance on Movielens-1M.}
    \label{fig:case_study}
\end{figure}

In this section, we present a specific example to illustrate the effectiveness of LLM reasoning in the feature selection process. Figure~\ref{fig:case_study}(a) shows the Cramér's V index of various features in the MovieLens-1M dataset, representing the correlation between different features. The darker the blue, the stronger the correlation between the two features. We observe that ``title'' and ``movie\_id'' are highly correlated, as indicated by the red box in the figure. Additionally, Figure~\ref{fig:case_study}(b) displays the AUC performance on the MovieLens-1M dataset for the best-performing baseline LPFS, LPFS-t (which removes the title feature based on LPFS), and SELF. We find that removing the title feature results in a significant improvement for LPFS, indicating that the strong correlation between ``title'' and ``movie\_id'' indeed hinders model training. Finally, we examine the feature importance ranking from LPFS: [\underline{\textbf{``movie\_id''}}, ``user\_id'', \underline{\textbf{``title''}}, ``genres'', ``zip'', ``gender'', ``age'', ``occupation'', ``timestamp'']. LPFS assigns high importance to both ``movie\_id'' and ``title'', likely because the model optimization process treats these two features as roughly equally important, overlooking the issue of information redundancy between them. However, SELF, by leveraging LLM reasoning, effectively identifies this issue. Below is an analysis extracted from the LLM response during prompt iteration regarding the title feature: ``While the movie title might have some influence, it's likely highly correlated with movie\_id and thus doesn’t add significant independent information.'' Consequently, a more accurate and effective feature ranking is obtained: [``user\_id'', \underline{\textbf{``movie\_id''}}, ``genres'', ``age'', ``gender'', ``timestamp'', ``occupation'', ``zip'', \underline{\textbf{``title''}}]. This demonstrates the effectiveness of SELF in handling feature importance extraction.

\section{Related Work}

\subsection{Feature Selection for DRS.}
Feature selection, an approach aimed at identifying predictive features from the original feature set, has become a critical part of the machine learning pipeline~\cite{liu2019automating,9547816,fan2020autofs,li2025survey,jia2024fine} and data mining framework~\cite{jia2025georanker,zhang2025lsrp,jia2024bridging,jia2024g3,jia2023mill}. Traditional feature selection methods can generally be categorized into three types: 1) Filter Methods. These methods define specific criteria for evaluating the importance of features, such as the Chi-squared test~\cite{liu1995chi2} and mutual information~\cite{battiti1994using}. 2) Wrapper Methods. These methods typically employ heuristic algorithms and use black-box models to assess the effectiveness of feature subsets~\cite{kohavi1997wrappers}. 3) Embedded Methods. This type integrates feature selection within the predictive model, evaluating features along with the model optimization process~\cite{lasso,gbdt}. With the advancement of deep learning technologies, numerous related works have emerged in the field of recommender systems, leading to the development of various feature selection methods~\cite{li2024recent}. AutoField~\cite{wang2022autofield} was the first to introduce the use of AutoML for learning feature importance in recommender systems. It constructed an innovative controller network that learns importance weights for different feature fields. 
Recently, there have been some attempts~\cite{jeong2024llm} to use LLMs for feature selection, but the recommendation system context has not been considered.

\subsection{AutoML in Recommender Systems.}
In recommender systems~\cite{jia2018second,jia2024d3}, challenges such as feature selection, feature crossing, and embedding dimension design face the difficulty of large candidate sets and the need for specialized knowledge to optimize them. AutoML is a technology that effectively lowers the barrier for users by automating and simplifying the design and optimization of machine learning algorithms~\cite{he2021automl}. For feature selection, methods like AutoField~\cite{wang2022autofield}, AdaFS~\cite{lin2022adafs}, and OptFS~\cite{lyu2023optfs} have achieved significant improvements in model performance and efficiency by designing gating weights to automatically select features. In the area of feature crossing, AutoCross~\cite{luo2019autocross} uses beam search in a tree-based space to automatically select high-level cross-features. 

\section{Conclusion}

In this paper, we introduce SELF, an agency-light feature selection method for DRS. Specifically, we first propose a prompt iteration technique, which iteratively prompts LLMs to obtain an initial ranking of features in the semantic space. Then, we introduce the bridge vector, which connects the semantic information of features with the current recommendation task, further fine-tuning the feature importance ranking. We conduct extensive experiments on three public datasets, and the results demonstrate the superiority of SELF compared to other state-of-the-art methods. We also release our code online for ease of reproduction.

\begin{acks}
This research was partially supported by Hong Kong Research Grants Council's Research Impact Fund (No.R1015-23), Collaborative Research Fund (No.C1043-24GF), General Research Fund (No.11218325), Institute of Digital Medicine of City University of Hong Kong (No.9229503), and Huawei (Huawei Innovation Research Program).
\end{acks}
\appendix
\section{Appendix}

\subsection{Dataset Description} \label{sec:app_dataset}

In this section, we introduce the datasets we used in our paper. The statistics of datasets are illustrated in Table~\ref{tab:app_dataset}.

\begin{itemize}
    \item \textbf{Movielelns-1M.} The Movielens-1M dataset is a highly popular dataset in the field of recommender systems, with the task of predicting whether a user will like a particular movie. In this paper, we consider interactions with a rating greater than 3 as indicative of user preference, labeling them as positive samples. This dataset consists of 9 features, which allows for the evaluation of various feature selection methods under conditions with a limited number of features. In our study, we divide the dataset into training, validation, and test sets using a 7:2:1 ratio.
    \item \textbf{Aliccp.} The Aliccp (Alibaba Click and Conversion Prediction) dataset is derived from real-world data collected from the e-commerce platform Taobao. It contains over 80 million interaction records, encompassing more than 230,000 users and 467,000 items, making it a dataset that closely mirrors real-world scenarios. Additionally, it includes 23 features, enabling the evaluation of various feature selection methods under typical conditions. In our study, we follow the original partitioning strategy of the Aliccp dataset, using 50\% of the data for training, while the remaining data is split equally between the validation and test sets.
    \item \textbf{Kuairand.} KuaiRand is a dataset collected from the short video platform Kuaishou, based on real-world data. We have selected the ``pure'' version, which contains over 1 million interaction records. The primary reason for choosing this dataset is its provision of 96 features, making it well-suited for thoroughly evaluating the effectiveness of various feature selection methods in scenarios with a large number of features. We partitioned the dataset into training, validation, and test sets using a 7:2:1 ratio.
\end{itemize}

\subsection{More details on implementation} \label{sec:app_imp}
In this section, we will introduce more details on implementation to ease reproduction.
For SELF, we selected three LLMs: GPT-4, GPT-4o, and GPT-3.5. In the hyperparameter analysis, we also explored the performance when using four LLMs, with the additional LLM being Gemini-1.5-Pro. Our experiments were conducted on a single A100 GPU. The hyperparameter settings for other baselines are detailed in Table~\ref{tab:app_imp}.

\begin{table}
\centering
\caption{Hyperparameters for baselines.}
\label{tab:app_imp}
\resizebox{\linewidth}{!}{
\begin{tabular}{cc} 
\toprule
Method      & Hyperparameters                                               \\ 
\midrule
\multicolumn{2}{c}{\textit{Shallow Feature Selection Methods}}              \\ 
\midrule
Lasso       & alpha=1.0                                                     \\
GBDT        & oss=logloss, learning\_rate=0.1, n\_estimators=100            \\
RF          & n\_estimators=100, criterion=gini                             \\
XGBoost     & booster=gbtree, gamma=0, max\_depth=6                         \\ 
\midrule
\multicolumn{2}{c}{\textit{Gate-based Feature Selection Methods}}           \\ 
\midrule
AutoField   & update frequency: 10                                          \\
AdaFs       & pretrained epoch: 0; update frequency: 4                      \\
OptFS       & gamma: 5000; epsilon: 0.1                                     \\
LPFS        & epsilon update frequency: 100; epsilon decay: 0.9978          \\ 
\midrule
\multicolumn{2}{c}{\textit{Sensitivity-based Feature Selection Methods}}    \\ 
\midrule
Permutation & n\_repeats=5                                                  \\
SFS         & num\_batch\_sampling: 100                                     \\
SHARK       & None                                                          \\
\bottomrule
\end{tabular}}
\end{table}

\section*{GenAI Usage Disclosure}
Generative AI tools were used only for spelling and grammar correction purposes. No content was generated or modified beyond these basic editing tasks.

\clearpage
\bibliographystyle{ACM-Reference-Format}
\balance
\bibliography{references}

\end{document}